\documentstyle[sprocl,epsfig]{article}

\def\Journal#1#2#3#4{{#1} {\bf #2}, #3 (#4)}

\def\be{\begin{equation}}
\def\ee{\end{equation}}
\def\bea{\begin{eqnarray}}
\def\eea{\end{eqnarray}}

\begin{document}


\title{GALACTIC WINDS IN IRREGULAR STARBURST GALAXIES}

\author{F. MATTEUCCI}

\address{Department of Astronomy,Trieste University, Via G.B. Tiepolo 11, 
\\ 34100 Trieste, Italy\\E-mail: matteucci@ts.astro.it} 

\author{S. RECCHI}

\address{Department of Astronomy,Trieste University, Via G.B. Tiepolo 11, 
\\ 34100 Trieste, Italy\\E-mail: recchi@ts.astro.it}


\maketitle\abstracts{In this paper we present some results concerning
the study of the development of galactic winds in blue compact
galaxies.  In particular, we model a situation very similar to that of
the galaxy IZw18, the most metal poor and unevolved galaxy known
locally.  To do that we compute the chemo-dynamical evolution of a
galaxy in the case of one istantaneous isolated starburst as well as
in the case of two successive instantaneous starbursts. We show that
in both cases a metal enriched wind develops and that the metals
produced by the type Ia SNe are lost more efficiently than those
produced by type II SNe. We also find that one single burst is able to
enrich chemically the surrounding region in few Myr. Both these
results are the effect of the assumed efficiency of energy transfer
from SNe to ISM and to the consideration of type Ia SNe in this kind
of problem.  The comparison with observed abundances of IZw18 suggests
that this galaxy is likely to have suffered two bursts in its life,
with the previous being less intense than the last one.}

\section{One and two bursts of star formation}
We assume a single instantaneous burst occurring at the center of a
gas-rich dwarf surrounded by a dark matter halo. The mass of stars
formed is $M_{*}=6 \cdot 10^{5} M_{\odot}$, the gas mass is
$M_{gas}=1.7 \cdot 10^{7} M_{\odot}$ and the mass of the dark halo is
$M_{dark}=6.5 \cdot 10^{8} M_{\odot}$. These parameters are chosen to
reproduce, for what possible, the characteristics of IZw18 (unevolved
local galaxy).  The galactic region extends for 700 pc in the vertical
(z) direction and for 1kpc in the radial direction (R).  In order to
study the chemo-dynamical evolution consequent to the burst, we adopt
a 2-D hydrocode coupled with detailed chemical yields from: type II
SNe, type Ia SNe and low and intermediate mass stars (see Recchi et
al. 2001 for more details about this model). We follow the evolution
of the abundances of H, He, C, N, O, Mg, Si, Fe in the gas. The
evolution is followed for 375 Myr since the burst.  We also consider a
case with a second burst occurring after 300 Myr from the first one
The mass in stars formed during the first burst is
$M_{*}=10^{5}M_{\odot}$, whereas the mass in stars formed in the
second burst is $M_{*}=5.8 \cdot 10^{5}M_{\odot}$.  The initial gas
mass and dark matter halo are the same as in the one-burst case.  The
simulation lasts 450 Myr in total since the first burst.  The SN
efficiencies of energy transfer into the ISM from SNe is the same in
both cases, in particular: for type II SNe we assume that the
efficiency is $\eta=0.03$ and for type Ia SNe $\eta=1$. This choice is
due to results from Bradamante et al. (1998) for type II SNe
indicating that the first SNe to explode lose a large fraction of
their initial energy by radiation due to the cold and dense ISM in
their surroundings, whereas for type Ia SNe the efficiency is maximum
due to the fact that they explode into an ISM already hot and rarefied
(see Recchi et al. 2001, for details).

\section{Dynamical Results}
We find that the starburst triggers indeed a galactic wind and the
metals leave the galaxy more easily than the unprocessed gas
confirming previous results (e.g. McLow and Ferrara, 1999).  We find
that SNe Ia eject their metals more efficiently than SNe II since they
inject all of their initial energy into the ISM.  This is a new result
relative to previous studies since it is the first time that type Ia
SNe are taken into account.  At variance with previous studies (see
e.g. Tenorio-Tagle 1996) we find that most of the metals are already
in the cold gas phase after 8-10 Myr from the beginning of the burst,
due to the fact that the superbubble created by the SNe does not break
immediately and thermal conduction can act efficiently.

\section{Chemical results}
We find that one single instantaneous burst, occurring in a primordial
gas (no metallicity), at an age of $\sim 31$ Myr can reasonably
reproduce the abundances measured in IZw18 (see figure 1). From this
one would conclude that perhaps this galaxy is experiencing its first
burst of SF, although one cannot exclude a previous burst which
enriched the gas no more than 1/50 Z$_{\odot}$.  However,
as evident in figure 1, the correct N/O ratio would last only for a
very short time since for $t > 31$ Myr the N/O ratio will start to
increase outside the permitted observational range.  In addition,
Color Magnitude Diagram studies (e.g. Aloisi et al. 1999) indicate the
presence of an old underlying stellar population in IZw18, thus
suggesting that the two-burst case is more realistic.  An interesting
result is that the [$\alpha$/Fe] ratios in the gas outside the
galaxy (i.e. in the galactic wind) are lower than inside, due to
the larger ejection efficiency by type Ia SNe (more iron is lost than
$\alpha$-elements). This creates an interesting dichotomy which can
have important consequences on the evolution of the ICM, especially if
this effect will be found also in elliptical galaxies.

In the two-burst case, at the onset of the second burst the
metallicity in the burst region is $ \sim Z_{\odot}/50$, and a very
good agreement with the measured abundances in IZw18 is obtained in
this case, especially for the C/O and N/O ratios and for a larger
range of burst ages (see figure 1).  Our results suggest that a first
weak burst occurred more than 300 Myr ago and followed by a more
intense one with an age between 20 and 70 Myr can reproduce most of
the properties of this galaxy.  The [$\alpha$/Fe] ratios in the two
burst case are always lower outside than inside the galaxy, but
the effect is less evident than in the one-burst case.

\begin{figure}
\centering
\epsfig{file=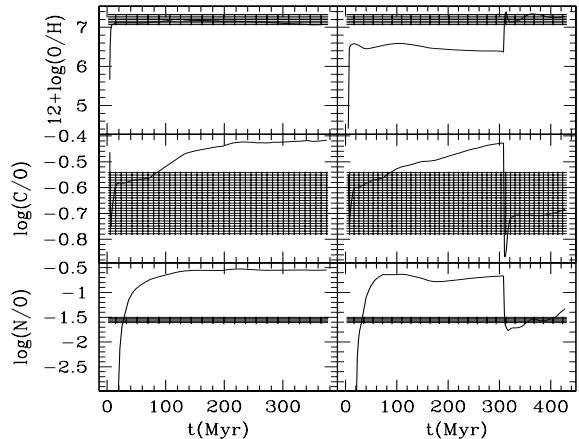, height=7cm,width=8cm}
\vspace{-1.cm}
\caption[]{\label{fig:fig 1} Evolution of O, C, N for the single-burst
model (left panels) and the double-burst model (right panels). The
superimposed grids represent the observative values found in literature
for IZw18}
\end{figure}

\section*{Acknowledgments}
We are gratefull to Annibale D'Ercole and Monica Tosi for their
help in this work.


\section*{References}
\begin{thebibliography}{99}
\bibitem{ma}Aloisi A., Tosi M. and Greggio L., \Journal{AJ}{118}{302}{1999}.

\bibitem{bu}Bradamante F., Matteucci F. and D'Ercole A., \Journal{A\&A}{337}
{338}{1998}.

\bibitem{bd}Mac Low M.-M. and Ferrara A., \Journal{ApJ}{513}{142}{1999}.

\bibitem{io}Recchi S., Matteucci F. and D'Ercole A., MNRAS in press (2001).

\bibitem{tt}Tenorio-Tagle G., \Journal{AJ}{111}{1641}{1996}

\end{thebibliography}
\end{document}